# The right time to learn: mechanisms and optimization of spaced learning

Paul Smolen, Yili Zhang and John H. Byrne

Department of Neurobiology and Anatomy, W. M. Keck Center for the Neurobiology of Learning and Memory, The University of Texas Medical School at Houston, P.O. Box 20708, Houston, Texas 77030, USA

Abstract | For many types of learning, spaced training that involves repeated long inter-trial intervals (ITIs) leads to more robust memory formation than does massed training that involves short or no intervals. Several cognitive theories have been proposed to explain this superiority, but only recently has data begun to delineate the underlying cellular and molecular mechanisms of spaced training. We review these theories and data here. Computational models of the implicated signaling cascades have predicted that spaced training with irregular ITIs can enhance learning. This strategy of using models to predict optimal spaced training protocols, combined with pharmacotherapy, suggests novel ways to rescue impaired synaptic plasticity and learning.

Repetitive training helps form a long-term memory. Training or learning that includes long intervals between training sessions is termed spaced training or spaced learning. Such training has been known since the seminal work of Ebbinghaus to be superior to training that includes short inter-trial intervals (massed training or massed learning) in terms of its ability to promote memory formation. . Ebbinghaus stated[1]: "…*with any considerable number of repetitions a suitable distribution of them over a space of time is decidedly more advantageous than the massing of them at a single time.*" His studies were based on the self-testing of acquired memory for lists of syllables, but the superiority of spaced training has now been established for many additional forms of human learning. For example, spaced learning is more effective than massed learning for facts, concepts and lists[2, 3, 4], skill learning and motor learning[5, 6], in classroom education (including science learning and vocabulary learning)[7, 8, 9], and in generalization of conceptual knowledge in children[10]. Spaced training also leads to improved memory in invertebrates, such as the mollusk *Aplysia californica*[11, 12, 13, 14], *Drosophila melanogaster*[15, 16] and bees[17], and in rodents[18, 19] and non-human primates[20, 21]. Memory extinction is commonly considered to involve the formation of a new memory, and in rat fear conditioning, spaced extinction trials are more effective than massed trails at establishing new memories[22].

Although it has been established that spaced training is superior to massed training in terms of inducing memory formation, key questions remain. What are the mechanisms underlying this superiority? Is it possible to use this mechanistic information to determine the optimal intervals between learning trials? If so, are fixed, expanding, or irregularly spaced intervals optimal? Another key question is whether an understanding of the mechanisms for optimal intervals can provide insights into the design of pharmacological approaches for memory enhancement. Computational models based on such a mechanistic understanding may be able to predict more complex approaches to memory improvement in which the application of multiple drugs, or combinations of drugs and training protocols, can enhance memory or treat deficits in learning and memory.

In this Review, we describe how new insights from molecular studies may help explain the effectiveness of spaced training, and how the molecular findings relate to the traditional learning theories that aim to account for this effectiveness. We also review how models of signaling pathways that are involved in synaptic plasticity can suggest, and experiments empirically validate, training protocols that improve learning and that rescue plasticity impaired by deficits of key molecular components. Finally, we discuss recent models that have suggested combined-drug therapies that may further enhance some forms of learning and that may have synergistic effects with optimized spaced learning on memory formation.

**Traditional learning theories**

We briefly summarize three of the well-known cognitive theories proposed to explain spaced training's superiority to massed training: encoding variability theory, study-phase retrieval theory and deficient-processing theory.

Encoding variability theory[23, 24, 25] posits that repeated stimulus presentations or learning trials are more likely to occur in multiple contexts if they are spaced further apart in time, and that a memory trace for repeated trials therefore includes elements of each of these contexts. Thus, spaced training would tend to bind together more contexts and hence form a more robust memory, as a greater number of testing contexts could elicit retrieval of the memory.

Study-phase retrieval theory[26, 27, 28, 29] posits that spaced stimulus presentations or learning trials are more effective than massed trials in reinforcing memory because each spaced trial elicits retrieval and reactivation of a memory trace formed by the preceding trial. In contrast, with short massed trials, the preceding memory trace is still active, so it is not retrieved or reactivated. Therefore memory cannot be reinforced. Study-phase retrieval theory also accounts for a decline in learning with excessively long intervals because in that case the preceding memory trace can no longer be retrieved. A recent variant, retrieved context theory, also incorporates elements of encoding variability theory and has succeeded in predicting the results of subsequently performed spaced learning experiments in humans[30].

Deficient processing theory posits that spaced training forms a stronger memory than does massed training because in the latter, some processes that are necessary to form memories are not effectively executed. The reasoning here becomes clearer by examining variants of this theory that specify the nature of the deficient process. One variant posits that excess habituation during massed trials prevents effective reinforcement of memory traces[31], whereas others posit that there is a failure to consolidate a memory (also known as consolidation theory)[32, 33], a lack of voluntary attention to massed presentations[31, 34], or a lack of cognitive rehearsals or reactivations of a memory trace within the short intervals that are characteristic of massed training[27, 35].

Consolidation occurs as a memory trace becomes more fixed and stable with time after training[2, 36]. Thus, consolidation theory[37, 38, 39] posits that a long-term memory trace is more efficiently stabilized or strengthened by spaced trials. The lack of cognitive rehearsals variant of deficient-processing theory might also be considered a more specific form of consolidation theory, because it assumes that a minimum number of rehearsals, or autonomous reactivations, are required to consolidate a memory trace. Variants of deficient-processing theory and relevant experiments are further discussed in REFs[28, 40, 41]. Below, we focus substantially on consolidation theory because, of all the traditional learning theories, it seems to be most closely aligned with our current understanding of the cellular and molecular mechanisms of memory.

Landauer[37] was one of the first to develop a conceptual model of the ways in which consolidation principles could explain the effectiveness of spaced training. Although the model was originally developed to explain the effects of short spacing intervals on memory formation, it can readily be generalized for the effects of arbitrarily long intervals (FIG. 1). The model is based on two assumptions. First, the state of a neural circuit following the first learning trial is such that a second reinforcing trial soon after will not markedly increase the consolidation of the learning trace resulting from the first trial (FIG. 1a). Thus, in massed training, overlap between traces and, consequently, saturation of an unspecified molecular mechanism diminishes the traces' summed impact on the consolidation of memory. Only when the effects of the first trial decay can the effects of a second trial be fully expressed (FIG. 1b, 1c), leading to greater potential consolidation of a memory in spaced training than in massed training (greater net gain, FIG. 1d). The second assumption is that the probability that the second trial can successfully reinforce the first trial declines with time (FIG. 1e). Actual consolidation is the product of these two assumptions, yielding a prediction of an optimal interval for spaced learning (FIG. 1f). Peterson[38] described a similar model focused on the dynamics of verbal learning.

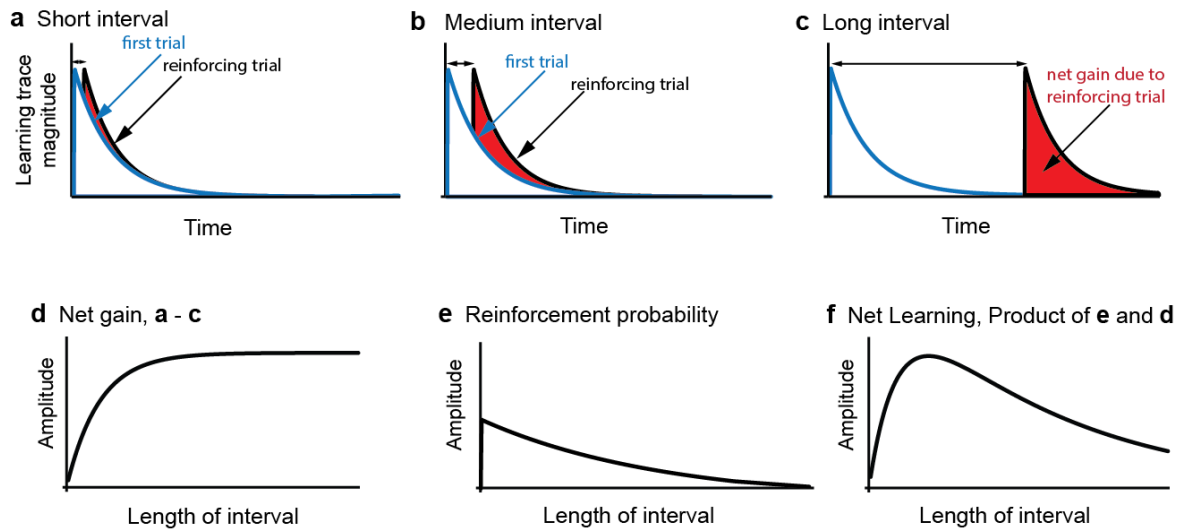

Figure 1 | **Early conceptual model of how learning trace dynamics generate an optimal interval**. **a–c** | As described by the early model of Landauer[37], spaced training is more effective at strengthening some form of trace corresponding to memory storage in the brain, although this conceptual model does not posit a biochemical or structural form for the trace. This model posits that memory formation becomes more effective with longer inter-stimulus intervals between training sessions because of decreasing temporal overlap between successive, short-lived learning traces. These learning traces do not themselves constitute a memory. However, their net effect, denoted here as net gain, contributes to the formation of a long-lived memory trace. In (**a–c**), learning traces elicited by two successive trials are shown. The model assumes that, for each value of the inter-trial interval length, a quantity denoted "net gain due to the reinforcing trial" is proportional to the red area. Shorter intervals are associated with more overlap of learning traces and less net gain. Therefore, a reinforcing trial is most effective after a refractory period following the preceding trial. For this conceptual theory, units for amplitude and time are arbitrary. **d** | A greater summed effect, or net gain, of reinforcing trials occurs for longer inter-stimulus intervals. The effect reaches a plateau for long intervals as the overlap between successive learning traces goes to zero **e** | Over longer times, a different quantity — the probability that a reinforcing trial will be effective at all in reactivating processes that constituted the preceding learning trace — declines. **f** | An optimum interval for maximizing the strength of the long-lived memory trace results when the greater net gain of reinforcement at longer intervals (from **d**) is multiplied by the slowly declining probability that a reinforcement will reactivate a previous learning trace (from **e**). The optimum interval for net learning is the one that produces the peak level of the trace in **f**.

Wickelgren[39] extended consolidation theory by positing that the resistance of a memory trace to decay increases with the age of the trace over the total duration of a spaced learning protocol. Thus, a trace would become not only strong but also highly resistant to decay following spaced trials.

**Molecular traces of time**

Substantial progress has been made in understanding the molecular mechanisms of memory. Given this progress, in this section we focus on potential molecular mechanisms of the spacing effect on long-term memory formation.

There is now agreement that learning is implemented, at least in part, by changes in synaptic strength (synaptic plasticity). For example, fear-conditioned memories can be alternately erased and re-instantiated by long-term depression (LTD) and long-term potentiation (LTP), respectively, of a defined synaptic pathway[42]. Thus, the molecular processes that are essential for spaced learning might reinforce extant LTP.

Reliable correlates of LTP are the remodeling and enlargement of postsynaptic dendritic spines, which are small protrusions that are associated with most excitatory synapses[43]. Thus, studying the differential dynamics of dendritic spine remodeling following massed versus spaced stimuli is likely to give insight into processes underlying the effectiveness of spaced training. Studies using rat hippocampal slice found that LTP induced by multiple trains of theta-burst stimuli was accompanied by extensive remodeling of synaptic ultrastructures[44, 45] and that subsequent spaced trains of theta-burst stimuli with intervals of 60 min or more between trains, were needed for optimal reinforcement of LTP[46]. Stimulated dendritic spines were remodeled over > 1 h, leading to enlargement of the existing functional postsynaptic density[45] and the presynaptic active zone[44]. The resulting increase in the numbers of AMPA-type and NMDA-type receptors at the synapse correlated with the magnitude of LTP.

Two hypotheses that involve spine remodeling have been put forward to explain the greater efficacy of spaced trials over massed trials in memory formation. These hypotheses have a common theme, which is that the learning process includes a refractory period during which the second of two closely spaced stimuli would be ineffective in enhancing the effects of the first (FIG. 2a). One hypothesis is that spaced but not massed repetitions of a stimulus allow the refractory period to be overcome and lead to repeated enlargement of a set of spines and a strengthening of the synaptic connections mediated by these spines[47]. (FIG. 2b) A second, not mutually exclusive, hypothesis[47, 48] is that molecular processes enable later spaced stimuli to induce LTP at spines that do not undergo initial enlargement. In this case spaced, but not massed, inter-trial intervals would allow for a molecular process termed 'priming' to be completed at these additional spines. After being primed, these spines would be strengthened by subsequent stimuli and incorporated into the memory trace (FIG. 2c). Currently, the molecular components of such a priming process are not known.

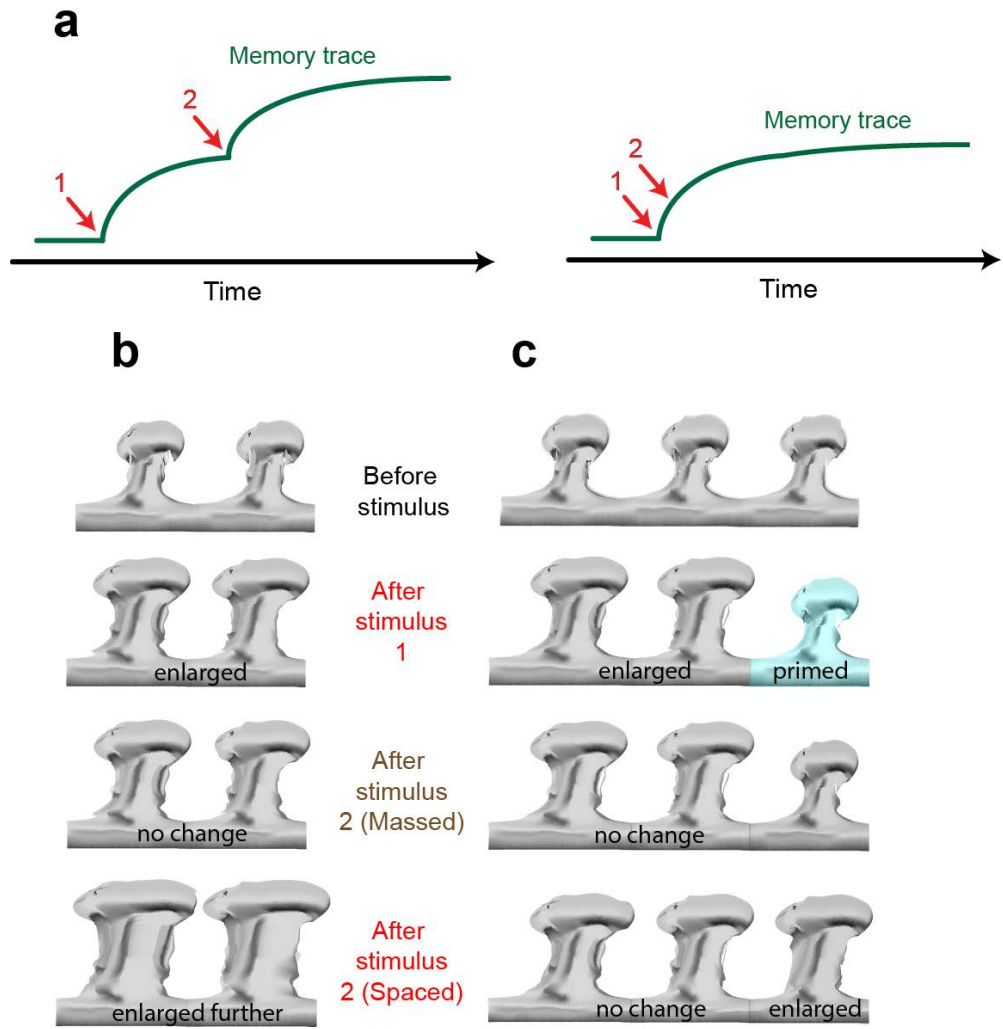

Figure 2 | **Model and hypotheses describing synaptic strengthening during spaced learning**. **a** | In the refractory-state model, spaced stimuli (left panel, stimulus 1, red arrow 1, followed substantially later by stimulus 2, red arrow 2) cumulatively strengthen a memory trace (as indicated by green line). By contrast, massed stimuli (right panel, stimulus 1 followed shortly after by stimulus 2) fail to cumulatively strengthen the memory trace. **b** | The cumulative synaptic strengthening in spaced training may be due to progressive enhancement of long-term potentiation, which could result from successive increases in the in the strength of the same synaptic contacts (shown here as successive increases in the volume of the same postsynaptic dendritic spine). Thus in one of two current hypotheses describing synaptic strengthening during spaced learning, stimulus 1 enlarges a population of spines. If stimulus 2 follows shortly after the first stimulus (as in massed training), it cannot further affect spines. However, if stimulus 2 comes after a refractory period (as in spaced training), it can further enlarge the same population of spines. **c** | Alternatively, enhancement of LTP could result from successive rounds of strengthening of new synaptic contacts. Thus in the second current hypothesis, stimulus 1 only enlarges a subset of affected spines, but primes additional spines. If stimulus 2 follows shortly after stimulus 1 (as in massed training), it has no effect. If stimulus 2 comes later (as in spaced training), it does not further enlarge the first subset of spines. Instead, stimulus 2 enlarges those spines that were primed, but not enlarged, by stimulus 1.

Through use of Schaffer–commissural projections in rat hippocampal slices, two studies[47, 48] have characterized the recruitment of additional synaptic contacts with the application of spaced stimuli. Theta burst stimuli applied at intervals of 10 min or 40 min did not cumulatively increase LTP. However, for longer intervals (60 or 90 min), a cumulative increase in LTP was observed over three bursts of stimulation. Each theta burst stimulus (TBS) led to actin filament polymerization in spines, which is known to be important for the stabilization of LTP[49]. The second stimulus (TBS2) yielded polymerization in spines that were not apparently affected by the first stimulus (TBS1), if TBS2 followed TBS1 by 60 or 90 min. These data do not suggest that successive theta burst stimuli further strengthen the efficacy of the same spines. Instead, they suggest that TBS1 initiates priming at all synaptic contacts of the stimulated afferents but only initiates consolidation and strengthening at a subset of contacts. Spines that undergo priming but not consolidation exhibit a refractory period of ~ 60 min, suggesting that priming takes time to complete (FIG. 2a). If TBS2 is given after the refractory period, some or all of the primed spines undergo consolidation. These data are consistent with the second hypothesis presented in the preceding paragraph, because TBS1 appears to enlarge and strengthen some spines but at others, TBS1 only initiates priming. These primed spines can then be strengthened by TBS2.

The dynamic properties of transcription factors and their interactions could also account for the superior efficacy of spaced training. LTP that persists for several hours or more requires translation and transcription[50, 51], which is reliant on key transcription factors such as cAMP response element (CRE) binding protein (CREB)[52]. Spaced training may be more effective, in part, because it may allow transcription factors such as CREB sufficient time to be activated, bind to promoters and induce a round of transcription for the consolidation of LTP[53] or long-term synaptic facilitation (LTF)[54]. In massed training, the trials would come too close together to initiate separate rounds of transcription. Indeed, in co-cultures of sensory and motor neurons from *Aplysia*, five spaced serotonin (5-HT) applications, each of 5 min with an inter-stimulus interval of 20 min (an analogue of spaced training) robustly elicits LTF that lasts for over 24 h[14], whereas 5-HT applied continuously over 25 min (an analogue of massed training) fails to yield reliable LTF.

In these sensory neurons, levels of a transcription activator CREB1 are elevated for at least 24 h after the spaced 5-HT treatments[54, 55]. This prolonged elevation of CREB1 owes to a positive feedback loop in which CREB1, acting via binding to a CRE regulatory element near *creb1*, increases the expression of *creb1*[54, 55] and other genes upregulated by CREB1. In addition, in these sensory neurons, the level of the transcription repressor CREB2 shows a late drop at ~12 h after treatment[56]. This drop in CREB2, coupled with the rise in CREB1, plausibly corresponds to an increased potential for gene induction. Thus, an additional 5-HT pulse near 12 h after treatment might optimally reinforce LTF.

LTF at these sensorimotor synapses is associated with a simple form of learning, long-term sensitization (LTS) of withdrawal reflexes. *In vivo*, four spaced electrical stimuli (30 min intervals between the stimuli) yield LTS that lasts > 24 h, with a weak residual of LTS being detectable at 4 d, and repetition of this spaced protocol once per day for 4 days yields much stronger LTS that lasts for over 1 week[13, 57, 58]. These data suggest that in this system, the dynamics of transcription activation and gene expression have slow components that can summate over multiple days, yielding long-lasting memory.

Recent data also illustrates that in the hippocampus, CREB and CCAAT enhancer binding protein (C/EBP), another transcription factor that is important for LTP, can remain active for many hours after learning. Following inhibitory avoidance training in rats, late peaks in brain-derived neurotrophic factor (BDNF) expression and C/EBP expression occur at ~12 h post-training, and inhibiting BDNF action at this time blocks memory maintenance[59]. These BDNF dynamics result from a positive feedback loop in which *C/ebp* induction leads to *Bdnf* upregulation, with the resulting increase in BDNF further activating the C/EBP signaling pathway[60]. Although this slow feedback loop was activated by single-trial training rather than spaced training, it would be of interest to model these dynamics, and to examine whether an additional spaced trial at ~ 12 h post-training, leading to a second induction of C/EBP at the time of elevated C/EBP, might optimally reinforce learning. A second prediction would be that massed stimuli are less effective if repeated at an interval too brief to allow the transcription regulation and *Bdnf* expression necessary to activate this feedback loop. Insights that can be obtained from computational models of learning are discussed later in the article.

On a shorter time scale, the dynamics of second messengers, kinases and phosphatases may contribute to the superiority of spaced training. One study with mice[61] found marked phosphorylation and activation of CREB in the hippocampus and cortex when object recognition trials were separated by an interval of 15 min but not by an interval of 5 min. Protein phosphatase 1 (PP1) appeared to be necessary for this spacing effect, because PP1 inhibition allowed the shorter interval to activate CREB. A study involving *Aplysia* sensory neuron–motor neuron co-cultures[62] found that protein kinase C (PKC) is activated to a greater extent during a massed stimulus (continuous 5-HT application) than a spaced stimulus (15 min intervals between applications). It is known that PKC acts to downregulate protein kinase A (PKA) and that PKA activation is necessary for LTF, thus these data delineate crosstalk between signaling pathways such that LTF is suppressed, in part, by stronger PKC activation during massed training.

Another study[16] characterized the dynamics of MAP kinase (MAPK) and of MAPK phosphatase in *Drosophila*. In an olfactory learning protocol, each spaced training trial generated a distinct wave of MAPK activity, whereas massed training trials occurred too close together to generate distinct waves. The authors therefore hypothesized that effective learning depended on the generation of distinct waves, which was only seen after the spaced trials.

Another phosphorylation-based mechanism has also been hypothesized to help explain the efficacy of spaced intervals in *Drosophila*. Spaced (15 min) intervals were more effective than massed (1 min) intervals in inducing olfactory learning, even given the same total training time (and thus more massed presentations)[63]. Two isoforms of *Drosophila* CREB — dCREB2-a and dCREB2-r — can activate and repress transcription, respectively. The authors proposed[64] that the kinetics of the phosphorylation of these isoforms differed such that the kinase activation generated by less frequent, spaced trials was sufficient to phosphorylate and activate dCREB2-a, whereas dCREB2-r could only be effectively phosphorylated by massed trials. Thus, training involving spaced intervals could maximally activate transcription and possibly induce the formation of long-term memory by activating dCREB2-a but not the counteracting repressor dCREB2-r.

Computational simulations have supported the plausibility of this mechanism[65], but it has not been validated empirically. However, it appears likely that a similar type of mechanism that is based on competition between an activator of long-term memory formation and a repressor, with the repressor only activated at short intervals, might be needed to explain any similar data in which massed training is less effective than space training even given equal total training times.

In experiments with *Aplysia*, when two electric shocks were given to induce LTS, maximal LTS was produced when the inter-stimulus interval was 45 min. LTS was not produced with intervals of 15 or 60 min[66]. The 45 min optimum was associated with activation of MAPK. Following either a single 5-HT pulse or a single shock, MAPK activation peaked at or near 45 min post-trial[12, 66], thus a 45 min interval might optimally reinforce the effects of MAPK. It is known this delayed MAPK activation requires protein synthesis[12], although the upstream mechanisms underlying the dynamics of the peak in MAPK activity at ~ 45 min are not well understood. Nevertheless, the key finding from these studies is that delayed activation of MAPK is intimately associated with the effectiveness of spaced stimuli to induce long-term memory.

Similarly, training with intervals of 60 min, but not 20 or 120 min, enhanced object recognition learning in wild-type mice and in a mouse model of fragile X syndrome model (*FMR1* knockout mice), at least partly by increasing synaptic activation of extracellular signal regulated kinase (ERK1 and ERK2: isoforms of MAPK)[67]. This 60 min interval was predicted to be optimal for learning because stimuli separated by 60 min had previously been found to enhance LTP in wild-type rodents[47]. Thus, in *Aplysia, Drosophila*, and mammals, MAPK activation appears to be a component of the molecular mechanism that underlies the spacing effect.

Some of these molecular mechanisms appear to fit with a theory in which spaced training sessions are effective because they reinforce the same memory trace or group of strengthened synapses. However, spaced stimuli might also reinforce memory by recruiting new synapses. ERK1 and ERK2 (ERK1/2) activation is needed for some forms of LTP[68] and one study[69] compared ERK1/2 activation in rat hippocampal pyramidal neurons following three spaced tetanic bursts (at 5 min intervals) with that after three massed bursts (at 20 s intervals). About twice as many dendrites with active ERK1/2 were found following spaced bursts, suggesting that spaced trials may recruit additional synapses on different dendrites for LTP. Thus, a range of molecular and cellular mechanisms appears to contribute to the efficacy of spaced training, in parallel or in series.

An extremely broad range of inter-trial intervals ranging from seconds to days has been used for spaced training (FIG. 3). For example, in honeybee olfactory learning, efficient spaced training can occur with intervals as short as 1 min[17]. Such brief intervals might allow for the reinforcement of the activity of a short-lived second messenger such as cAMP that is produced by preceding trials. The dynamics of kinase activation constitute a second substrate of spacing effects. In *Aplysia, Drosophila,* and mammals, the data discussed above indicate that commonly reported intervals, ranging from ~ 5 min to 1 h, may allow for the reinforcement of the activities of key kinases essential for LTP or LTF, and consolidate structural changes in dendritic spines.

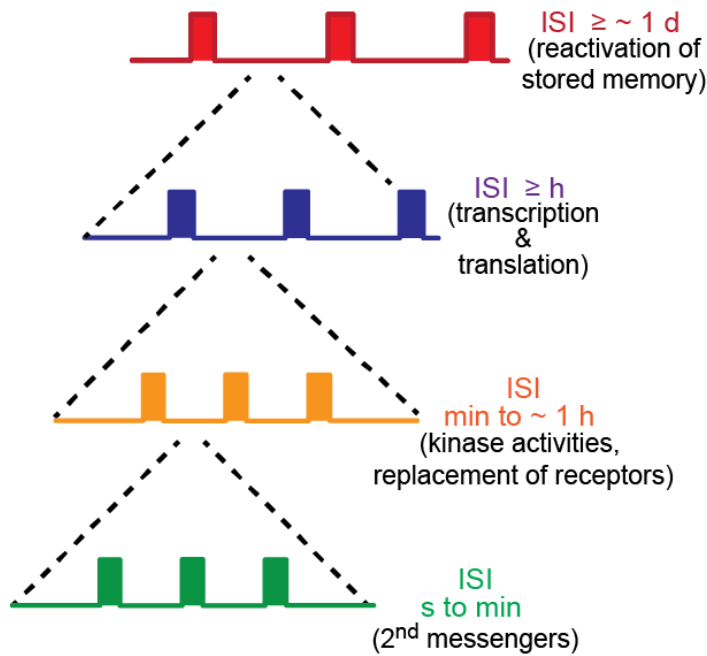

Figure 3 | **Different mechanisms may underlie enhancement of learning by spaced intervals of widely varying lengths**. For relatively brief inter-trial intervals (bottom trace), successive trials may coincide with and reinforce peak second messenger levels generated by preceding trials. In each trace, individual rectangles represent individual trials, and converging lines between traces represent the lengthening of time scales as one moves upwards in the illustration. For somewhat longer inter-trial intervals (several min to ~1 h), successive trials may reinforce the peak activities of kinases elicited by preceding trials and elicit LTP of primed dendritic spines. Intervals of this length may also, in hippocampus, be needed to allow replacement of inactivated receptors at stimulated spines[82], enabling succeeding stimulus repetitions to potentiate those spines. For intervals of ~ 1 h or more, succeeding trials may also align with peaks in transcription factor activity and gene expression owing to preceding trials. For the longest inter-trial intervals (many h or longer) succeeding trials may reactivate and thereby further potentiate consolidated memory traces. All these processes are likely to contribute to the consolidation of long-term memory, in many if not all species. However, depending on the inter-trial interval length used in a particular spaced learning protocol, the dynamics of a particular type of process (e.g. kinase activation) may contribute in particular to the efficacy of spaced learning. In addition, trials at one temporal domain (e.g., 1 day) may be unitary events, but also may constitute a block of spaced trials from another temporal domain (e.g., min to h). For example, an effective protocol for long-term sensitization training in *Aplysia* is the use of four trials with an intertrial interval of 30 min, with this block repeated four times with a one-day intertrial interval[13]. Thus, some effective training protocols consist of a hierarchy of temporal domains of training sessions, with briefer sessions embedded within longer ones. In this illustration intervals are shown with regular spacing, but more effective learning may occur with irregular spacing (see below).

It is plausible that the minimum inter-stimulus interval for effective learning, for a given protocol and system, corresponds to the interval that is necessary to allow each stimulus to contribute separately to a rate-limiting biochemical process. For example, for rapid honeybee olfactory learning with an effective interval of 1 min, the rate-limiting process might be second messenger accumulation or rapid activation of a kinase. For even shorter intervals, the time scale of the rate-limiting process might be too long to permit each brief stimulus to contribute separately to the process — a group of closely spaced stimuli would instead tend to act as just a single stimulus. For intervals of 1 min or more, each stimulus would be able to contribute a discrete increment to the rate-limiting process, allowing effective learning. For the spaced LTP protocol of Gall, Lynch and colleagues[47, 48], an interval of 40–60 min is needed for successive TBS to further increase LTP. Here, the rate-limiting process would be different — plausibly slower activation of an unspecified kinase or other intracellular signaling event, with a time constant near the minimum effective interval of ~ 40 min. Stimuli at intervals much shorter than this would not be able to generate summation of the rate-limiting process and would therefore not cause additional LTP.

For other systems, a similar assumption may apply to the dynamics of transcription activation. For LTF and LTS in *Aplysia*, transcription, as discussed above, may constitute a rate-limiting process that helps determine the efficacy of spaced training. However, it is evident that even for systems such as honeybee olfactory learning that involve short, spaced intervals, effective long-term memory formation relies on the activation of transcription and translation, downstream of the intracellular signaling pathways that are activated by these intervals[17, 70]. Reactivation of memory traces may constitute an additional temporal substrate that underlies the longest reported effective intervals, on the order of a week[71]. Such intervals are likely to reactivate and reinforce consolidated patterns of strengthened synapses that correspond to memory traces that are maintained by neuronal network activity[72]. Spaced learning with these long intervals would reactivate critical components at these synapses, and in particular reactivate NMDA receptors at these synapses. Studies using inducible and reversible NMDA receptor knockouts have demonstrated that such NMDA receptor reactivation, which may also in part result from spontaneous neuronal activity, is required to sustain remote memory storage[73, 74]. Positive feedback loops that maintain key kinases and other molecules in persistently active states at strengthened synapses may also contribute to such long-term memory storage[75, 76, 77, 78, 79]. An important topic for future research will be to further investigate the molecular processes that support effective spaced learning for humans that involves inter-trial intervals of a day or more. An implication of Fig. 3 is that multiple temporal domains of spaced training can be engaged in spaced training. For example, an effective protocol for long-term sensitization training in *Aplysia* is the use of four trials with an inter-trial interval of 30 min, repeated four times with a one-day inter-trial interval[13]. Thus, at least in some cases, there appears to be a hierarchy of temporal domains of training protocols with briefer protocols embedded within longer ones.

The above considerations, and most empirical studies, are concerned with only typical, or minimum, inter-trial or inter-stimulus intervals for effective spaced learning or for the summation of LTP. Only a few studies have delineated, for any specific system (that is, a given species and stimulus protocol), both minimum and maximum effective intervals. One study[80] found that in a hippocampal slice preparation, 5–10 min intervals between tetani were ideal for induction of LTP and they produced similar levels of LTP, with longer or shorter intervals yielding both less LTP and less ERK1/2 activation. In *Aplysia,* LTS was effectively induced by an interval of 45 min between electrical stimuli, but not by intervals of 15 min or 60 min[66]. As noted above, the authors of this study hypothesized that the coincidence of peak MAPK activation with the second trial was necessary for effective learning. In addition, 60 min intervals were effective for forming object location memory in mice with three trials, but intervals of 20 min or 120 min were not[67].

Owing to the small number of such studies and the lack of sufficient characterization of the accompanying molecular processes, it is not yet possible to make detailed statements about the ways in which intracellular signaling pathways could co-operate to generate both minimum and maximum intervals. For maximum intervals, a reasonable qualitative assumption is that each trial or stimulus generates a separate, relatively short-lived biochemical trace, and that for effective spaced learning, these traces must overlap and summate, with the summed magnitude driving the formation of long-lasting synaptic potentiation. These dynamics would be analogous to the necessary overlap of traces in the conceptual model of Landauer (FIG. 1a-c). For intervals longer than the maximum, the individual biochemical traces would decay and not overlap.

**Recent data and learning theories**

Do the biochemical and morphological mechanisms proposed to contribute to the greater efficacy of spaced training align with traditional cognitive theories? At this point, much of the extant cellular data seem to be compatible with the deficient processing theory, particularly two of its variants: the consolidation theory and the lack of cognitive rehearsals theory. In the consolidation theory, intervals between massed trials are proposed to be too short for the consolidation and consequent summation of memory traces that are engendered by successive trials. In the cognitive rehearsals theory, massed trials are proposed to lead to fewer cognitive rehearsals, or autonomous reactivations, than do spaced trials, and therefore less cumulative consolidation and persistence of a memory.

The required refractory period of ~ 1 h between successive theta-burst stimuli in to induce progressive increments in hippocampal LTP[46, 47] may be in line with the first of these variants, that short intervals are insufficient for consolidation and consequent summation of memory traces. The refractory period appears to be necessary to complete the 'priming' of dendritic spines that were stimulated, but not potentiated, by the first theta-burst. Priming allows these spines to potentiate after the second burst, and thus constitutes a biochemical stimulus trace (FIG. 1d and FIG. 2a). Kramár *et al.*[47] noted that in hippocampal slices, additional potentiation can be induced up to four hours after induction of the first LTP increment[81]. The stimulus trace associated with priming may therefore take at least four hours to decay. Such a long trace lifetime might allow a broad temporal window for optimal training trials.

In rat hippocampal slices, theta burst stimuli lead to proteolytic inactivation of integrin receptors at stimulated dendritic spines[82]. These receptors are then replaced by vesicular transport of new receptors over ~ 40 – 60 min, and it is hypothesized[82] that subsequent theta burst stimuli at these synaptic contacts cannot induce spine enlargement or LTP until after this replacement has occurred, thus accounting for the refractory period of ~ 1 h in order for a second theta burst stimulus to yield additional LTP. This receptor replacement may constitute, at least in part, the priming of dendritic spines discussed above. These hypothesized dynamics may be in line with deficient processing theory, with receptor replacement being the necessary process that can only occur during spaced inter-trial intervals (FIG. 3).

Transcription factor activation also constitutes a biochemical trace, and in some systems, training may only be effective if inter-trial intervals are long enough so that each trial can induce a separate round of transcription and translation. Similarly, short (massed) inter-trial intervals may not lead to sufficient levels, or a sufficient duration, of activated MAPK or other kinases to support the consolidation of long-term memory.

The variant of deficient processing theory positing that only spaced trials can generate sufficient cognitive rehearsals or reactivations of a memory to support long-term memory consolidation may also correspond with the empirical finding that repeated theta-burst stimuli, spaced by ~ 1 h, can recruit additional dendritic spines by potentiating spines that were primed by preceding stimuli. A memory reactivation would be analogous to theta-burst stimulus in that both events would initiate priming and potentiation. It also appears plausible that repeated memory reactivations might induce further rounds of transcription of genes involved in LTP, such as *C/ebp* and other CREB-activated genes, supporting further consolidation of long-term memory.

To more strongly connect this variant of deficient-processing theory to recent cellular and molecular data, one must also assume that reactivations of a memory reactivate some of the same neurons and synapses that were activated in the original learning sessions. In that way, the rehearsals and learning trials would reinforce memory in the same way. This assumption appears plausible but requires further empirical investigation. Although finer-grain analyses are necessary, a study using functional MRI during verbal learning supports this assumption[83]. In this study, a specific brain region associated with rehearsal of verbal memory, the left frontal operculum, was activated more during spaced learning of paired-word associations than during massed learning. We note that these posited memory reactivations, on time scales of ~ 1 h or longer, are distinct from voluntary rehearsals of a memory on a short time scale (seconds or ~ 1 min). Substantial behavioral evidence suggests that this latter voluntary, short-term rehearsal is not essential for spaced learning[31, 34].

The remaining variants of deficient processing theory, which focus on habituation or on a lack of voluntary attention during massed presentations, do not appear to relate as readily to the current single-neuron data. These variants have also been argued to not readily accommodate certain verbal learning observations[2]. As regards encoding-variability theory, data on neuronal network dynamics, rather than single-neuron data, will be needed to determine to what extent the binding of contexts to memory occurs, which is required in this theory. Similar data will also be needed to assess whether the binding of memories of later trials to those of earlier trials occurs, which is required in study-phase retrieval theory. It will be important to re-assess all these competing spaced learning theories as more information becomes available on the dynamics of memory networks. Indeed, different theories may be more or less applicable to different memory systems.

**Irregular spacing can enhance learning**

Attempts to optimize the spacing effect have generally been based on trial-and-error approaches. Consequently, most, if not all, training protocols used in animal and human studies are probably not optimal. For almost all learning paradigms, the training intervals are fixed, although in one type of spaced training paradigm, the intervals between sessions progressively lengthen[2, 84]. However, a meta-analysis[2] and a text learning study[84] found no substantial evidence for the superiority of this approach in terms of promoting long-term memory formation.

It appears evident that at least part of the improvement in learning that is found with spaced training protocols can be explained by the dynamic relationships between the training trials and the underlying cellular and molecular mechanisms that are associated with memory formation (FIG. 3). But is the inverse possible? Can knowledge of the dynamics of the memory mechanisms be used to enhance memory processing by predicting optimal training protocols, possibly with irregular training intervals? One approach is to develop models of the biochemical cascades that underlie memory formation and use simulations to rapidly test the effectiveness of different training protocols[85]. In recent years, models have described the dynamics of the biochemical reactions that transduce stimuli into LTP[86, 87, 88]. These models have differential equations that simulate and predict the dynamics of the activities of key molecular species. Simulations have reproduced the dynamics of MAPK during LTP induction[80, 86, 88]. Models have also simulated the activity time courses of PKA, CAMKII, other key enzymes and downstream transcription factors during LTP induction[88, 89, 90]. Each signaling cascade in these models displays a characteristic activity time course, thus it is likely some irregular sequence of intervals would be predicted to maximize the induction of LTP. For example, subsequent trials that are delivered at times that coincide with kinase activity peaks might optimally reinforce learning.

One study from our laboratory developed a model describing the 5-HT-induced PKA and ERK signaling pathways essential for LTF in *Aplysia*[85]. In the model (FIG. 4a), the necessity of PKA and ERK activation for LTF was simply represented with a variable termed 'inducer'. The value of inducer was proportional to the product of PKA and MAPK activities. The amount of LTF and LTS was predicted to increase with an increase in the peak value of inducer. Ten thousand different protocols consisting of five trials that were separated by intervals of 0 to 45 min were simulated (FIG. 4b). The ability to simulate and predict the effects of so many protocols in a relatively short space of time represents a distinct advantage of computational studies over empirical studies. The simulations determined that of these protocols, a massed protocol (FIG. 4b) produced the lowest peak value of inducer, consistent with data that massed 5-HT application fails to produce LTF[14]. The 'best' protocol yielding the highest peak value for inducer, termed the 'enhanced' protocol (FIG. 4b), had irregular intervals. The protocol termed the 'standard' protocol (five 5-min pulses of 5-HT, with uniform inter-stimulus intervals of 20 min) (FIG. 4b), has been commonly used to induce LTF in empirical studies for ~30 years[91]. This standard protocol yielded an intermediate peak value of inducer and was predicted to have an intermediate effectiveness (FIG. 4c). These predictions were empirically validated. LTF and LTS produced by the enhanced protocol exceeded that from the standard protocol (FIG. 4d)[85]. An explanation for this enhancement of LTF, consistent with data[11, 12, 92], is as follows. In response to each 5-HT pulse, PKA activates rapidly and decays rapidly (FIG. 4c). MAPK activity rises and decays more slowly, not peaking until ~45 min after a pulse. The four initial pulses initiate a surge of MAPK activity, which peaks near the time of the last pulse. This last pulse activates PKA, so that a PKA peak is approximately coincident with peak MAPK activation, maximizing inducer and the predicted LTF.

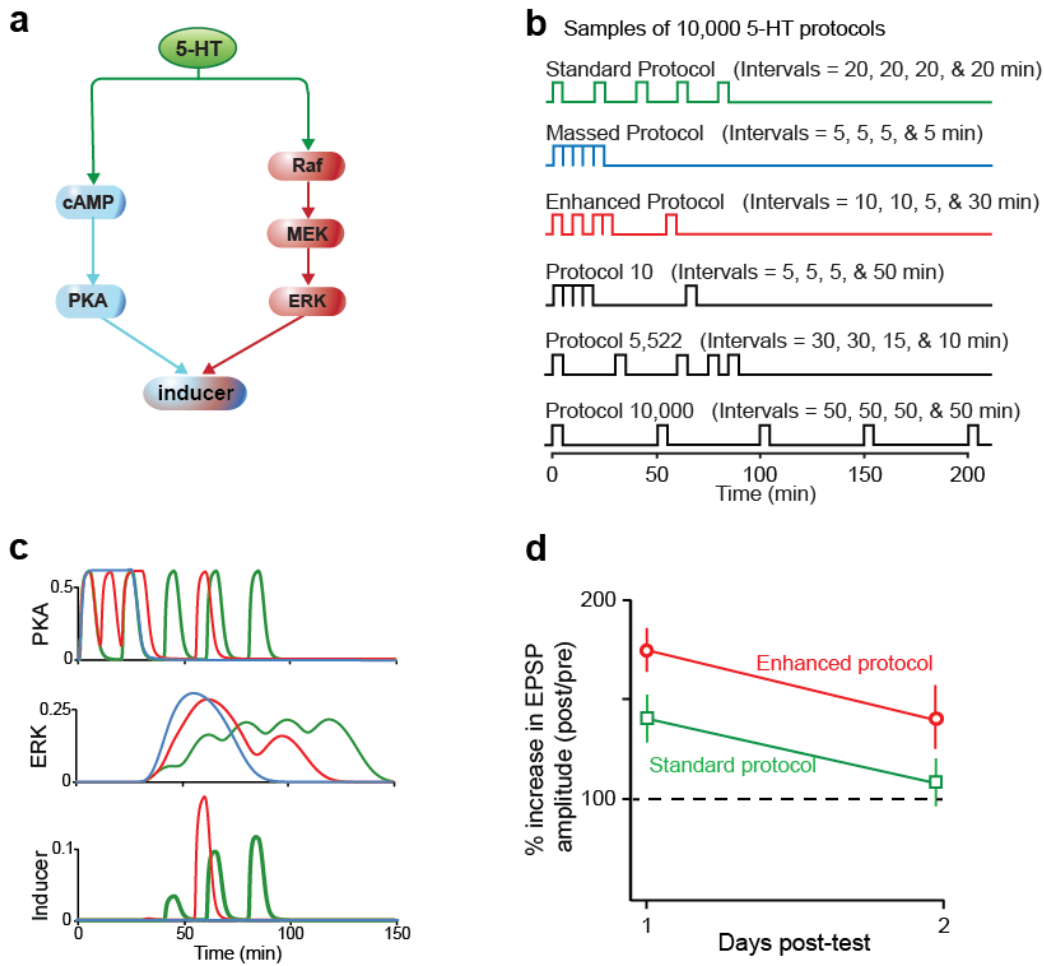

Figure 4 | **Dynamics of a model which has successfully predicted greater efficacy for a learning protocol with irregular spaced intervals. a |** A simplified mathematical model[85] describes the activation and effects of two key kinases necessary for LTF, a cellular correlate of a simple form of learning, long-term sensitization. Brief applications of 5-HT activate protein kinase A (PKA) via elevation of the secondary messenger cAMP, and activate the ERK isoform of MAP kinase via a Ras-Raf-MEK cascade. PKA and ERK interact, at least in part via phosphorylation of transcription factors, to induce LTF. In the model, the variable 'inducer' represents the PKA–ERK interaction. A higher peak value of inducer was assumed to predict a greater amplitude of LTF. **b |** Six samples of the 10,000 5-HT protocols that were simulated with the model. All protocols consist of five 5-min pulses of 5-HT, diagrammed as square

waves, with inter-pulse intervals chosen as multiples of 5 min, in the range from 5–50 min. The standard protocol (green trace) is the protocol most commonly used in studies of LTF *in vitro*. The enhanced protocol (red trace) produced the largest peak value of inducer, whereas the massed protocol (blue trace) produced the smallest peak value of inducer. The standard protocol has uniform inter-pulse intervals of 20 min, whereas the enhanced protocol has non-uniform intervals of 10, 10, 5 and 30 min. The massed protocol has no gaps between the 5-HT pulses. **c |** Simulated time courses of activated PKA, activated ERK and inducer in response to the standard protocol (green traces), enhanced protocol (red traces), and massed protocol (blue traces). The unit of concentration (y axes) is μM. **d |** In an empirical validation of the model's prediction, the LTF induced by the enhanced protocol, as determined by the percentage increase in the amplitude of excitatory postsynaptic potentials (EPSPs), was greater than the LTF produced by the standard protocol. Figure adapted from REF. 85.

If irregularly spaced protocols can enhance normal learning, might modeling also predict protocols capable of restoring learning that is impaired by a genetic mutation or other physiological insults? A recent study[90] tested this hypothesis. CREB binding protein (CBP) is an acetyltransferase and essential co-activator for several transcription factors including phosphorylated CREB (pCREB). CBP is also required for the consolidation of long-term memory[93]. Mutations that decrease CBP activity cause a human genetic disorder termed Rubinstein–Taybi syndrome (RTS)[94] that is associated with intellectual disability and learning deficits, and $Cbp^{+/-}$ mice show impaired LTP and LTM[95]. The recent study[90] used small interfering (siRNA) knockdown of CBP in *Aplysia* sensory neurons to impair LTF. In this study, the model previously used[85] to predict optimal, irregularly spaced protocols that would enhance LTF (FIG. 4a) was extended to represent induction of *c/ebp*, a transcription factor known to be essential for LTF[96]. In simulations of the effects of different spaced protocols, greater peak levels of phosphorylated C/EBP (pC/EBP) were taken to predict greater LTF. Simulations showed a substantial decrease in pC/EBP when CBP was reduced by a decrement that corresponded to the siRNA effect. A 'rescue' protocol with irregularly spaced intervals was predicted to restore peak pC/EBP and, correspondingly, LTF. This rescue protocol was empirically validated to restore normal LTF in *Aplysia*. A similar predicted rescue protocol of irregularly spaced intervals rescued a deficit in LTF that was produced by a siRNA knockdown of CREB1[97].

Although these empirical studies were conducted in *Aplysia*, it should be noted that key molecular mechanisms of memory are substantially conserved from simple model organisms such as *Aplysia* to mammals[52, 96]. For example, LTF and LTP both rely on PKA and ERK activation[12, 92, 98] and both rely on co-operative gene induction by pCREB and CBP[53, 96, 99, 100]. LTF relies on deactivation of CREB2, a transcriptional repressor[101]. Similarly, relief of transcriptional repression due to ATF4, a mammalian analogue of CREB2, appears to be important for the maintenance of hippocampal LTP[102, 103]. Thus, the results with *Aplysia* suggest it may be possible, in complex organisms including mammals, to computationally predict the efficacies of numerous learning or training protocols, a process that is impractical using empirical studies alone.

Given that knowledge of the underlying biochemical cascades can help develop models to predict optimal training protocols, can models also be used to predict pharmacological targets to improve memory? The time may also be right for such an approach. For example, if simulated LTP deficits were rescued by combined parameter changes corresponding to known drug effects, these 'best' parameter combinations might prioritize drug combinations for testing in animal models. A recent study[104] took a first step by modelling LTP induction and transcriptional regulation by CREB, and simulating the effects of drugs on LTP by altering the model's parameters. In this model, the magnitude of LTP induction was represented by an increase in a synaptic weight variable. LTP impairment seen in a mouse model of RTS[95] was first simulated. Then, starting from this simulation, the parameters were altered in ways corresponding to plausible single-drug effects. However, no single-drug effect completely rescued LTP. Therefore, pairs of parameter changes were considered, corresponding to plausible paired-drug effects. Two pairs were identified that restored LTP. In the first case, an increased rate constant for histone acetylation, corresponding to application of an acetyltransferase activator, was paired with an increased duration of stimulus-induced cAMP elevation, corresponding to application of a cAMP phosphodiesterase (PDE) inhibitor. The second pair corresponded to a PDE inhibitor paired with a deacetylase inhibitor. For both pairs, additive synergism, defined as a combined-drug effect that exceeds the summed effects of the individual drugs, was also evident, as quantified by a simple additive measure (FIG. 5). A subsequent empirical study by another group did find that pairing a PDE inhibitor with a deacetylase inhibitor was effective in rescuing a deficit of LTP in a mouse model of Alzheimer disease[105]. A further extension of these strategies might similarly predict, and empirically test, enhancement of synaptic plasticity when pharmacotherapy is combined with computationally designed spaced protocols.

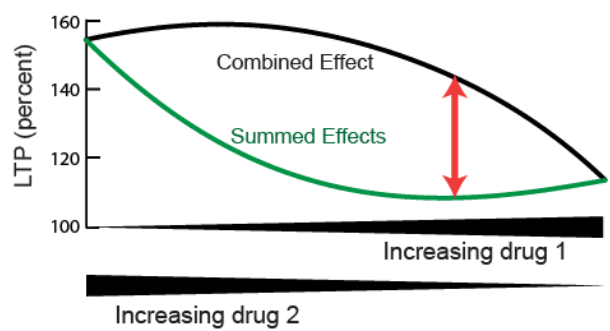

Figure 5 | **A model predicts that a pair of drugs can act synergistically to enhance LTP**. A CREB binding protein (*Cbp*) mutation impairs hippocampal LTP and impairs learning in mice, and *Cbp*$^{+/-}$ mice are considered a model for aspects of Rubinstein-

Taybi syndrome in humans[104]. We developed a model to examine whether drugs could be used to overcome this impairment in LTP. This figure was generated from a series of simulations of the effects of two drugs on the induction of LTP. LTP was modeled as the percent increase in a synaptic weight variable. In the absence of drugs, simulated LTP induced by a high-frequency tetanic stimulus was strongly impaired. Only a 50% increase in synaptic weight for C*bp*$^{+/-}$ occurred, compared to an increase in synaptic weight of 148% with non-mutated *Cbp*. The effect of each drug was simply modeled as a change in the value of a kinetic parameter. In this series of simulations, the doses of two drugs — drug 1, a cAMP phosphodiesterase inhibitor, and drug 2, an acetyltransferase activator — were concurrently varied. The effect of drug 1 was simulated by decreasing a rate constant for cAMP degradation, and the effect of drug 2 was simulated by increasing a rate constant for histone acetylation. The 'dose' of drug 1 — the amplitude of the rate constant change — was increased and simultaneously the dose of drug 2 was decreased. 80 pairs of drug doses were simulated. Both drugs substantially enhanced LTP. For drug 2 alone (left endpoint of graph) LTP was 155%, and for drug 1 alone (right endpoint), LTP was 116%. For both drugs together, with smaller doses of each drug, intermediate LTP amplitudes were observed (combined effect curve). This series of simulations further shows that additive synergism persists over a substantial range of drug doses. Additive synergism is quantified as the difference (red arrow) between the LTP simulated when both drugs are applied together (combined effect curve), and the LTP simulated by adding together the effect of the drugs applied individually in separate simulations (summed effects curve).

**Future directions**

There is reason for optimism that more predictive models for determining optimal intervals between learning trials will be available in the near future, because the molecular data that are necessary for the development of such models, delineating dynamics of signaling pathways that are important for LTP and long-term memory, continue to accumulate rapidly. However, despite the progress being made in understanding the molecular mechanisms of the spacing effect, some aspects of this effect cannot be explained by current models and constitute important directions for future research. For example, in human verbal learning, an interesting positive correlation exists between the length of inter-trial intervals for effective spaced learning and the retention interval (that is, the interval between the final training trial and the test of memory retention). With relatively short retention intervals (~1 min–2 h), training intervals in the broad range of ~1 min to 3 h yield greater verbal learning than do training intervals of 2 d or more[2]. With a longer retention interval of 1 d, a 1 d training interval yielded greater learning than did a very short (< 30 s) interval. For verbal learning with a retention interval of 6 months, a training interval of 7 d was superior to an interval of 3 d[71]. This correlation between longer training and retention intervals suggests that longer training intervals preferentially form a memory trace with a very long lifetime. For the temporal range of minutes versus hours, it is plausible that a longer trace lifetime corresponds, at least in part, to increased activation of transcription by the longer training intervals. However, this explanation may not suffice when comparing training intervals of ~ 1 d versus many days. It would be of interest to determine whether reactivation of stored memory representations at the network level, or transfer of these representations between brain regions, contribute to this correlation.

Another challenge will be to use innovative strategies to test the predictions of the cognitive theories for the spacing effect. For example, consider the variant of deficient-processing theory positing that repeated cognitive rehearsals of a memory are needed for consolidation. A neuronal correlate of rehearsals is, plausibly, repeated activation of a specific neuron assembly that serves as a locus of storage of a long-term memory trace. Empirically, is such repeated activation necessary for persistence of memory for days or longer? Repeated spontaneous activation of neuron assemblies does occur[106, 107] as does repeated replay or rehearsal of assemblies that encode recent experiences[108, 109]. One study supporting the necessity of such replay found that the post-training suppression of activity of neurons that were engineered to overexpress CREB in the amygdala for several h after conditioning blocked the consolidation of a memory of association between cocaine and a location[110]. Similar blocking effects were obtained by the indiscriminate activation of neurons that overexpressed CREB. Although encouraging, these manipulations lack the cellular precision that is necessary to demonstrate conclusively that reactivation of a particular assembly of neurons is essential for the persistence of long-term memory. Future studies using optogenetic techniques could provide that precision. Similarly, innovative strategies will be needed to address whether effective spaced learning requires the binding of contextual and episodic memories at the neuronal network level, such as posited by encoding variability theory, or increased binding due to greater retrieval effort, as posited by study-phase retrieval theory.

The successful prediction of the interval structure of behavioral training protocols that may overcome some human learning deficits (when applied alone or in combination with pharmacotherapy) will require improved knowledge of the signaling pathways that underlie LTP and long-term memory and of the ways in which the deficits affect those pathways. Future models are still likely to be incomplete owing to gaps in knowledge. For example, data will be incomplete and associated with unavoidable uncertainties in the values of biochemical parameters in models such as enzyme activities or protein concentrations. In model development, data from several preparation types (for example, cell cultures and slices) and species (for example, primates and rodents) commonly need to be used to estimate different parameters[86, 88]. However, although these limitations are important, the potential benefits of combining modeling with experiments in the ways discussed herein are extensive, such that this strategy may have promise for improving the clinical and educational outcomes for patients with learning and memory deficits. In addition, it is possible that education and learning in individuals without such deficits could benefit from such a strategy. Indeed, enhancing normal learning by judicious pharmacotherapy has recently received attention[111], and combining drugs with optimized spaced learning protocols might yield even better outcomes.

**Glossary**

Reinforcement: A broad term used here to describe a stimulus or item that enhances the strength or lifetime of a memory. Reinforcement stimuli activate biochemical and molecular processes and thereby regulate changes in synaptic strength associated with memory.

Memory Extinction: The decline of a learned behavioral response to a conditioned stimulus following the withdrawal of reinforcement stimuli that were previously paired with repetitions of the conditioned stimulus. This term is distinct from forgetting, which is the decline of a learned response during the prolonged absence of stimulus repetitions.

Habituation: A decrease of the behavioral response to a stimulus following frequent repetitions of that stimulus. This term is distinct from extinction, because habituation can denote a decrease of response to a stimulus that was never paired with a reinforcing stimulus.

Memory Reactivation: Re-instantiation of a conditioned behavioral response or of neural activity associated with a specific response. Reactivation can be elicited by presentation of a conditioning stimulus or of the context in which learning previously occurred, or it can be spontaneous, occurring as a part of normal ongoing neural activity.

Drug Synergism: In combined-drug treatment, a synergistic effect of the combination is an effect greater than what would be predicted by considering the individual drugs as independent and not interacting. Several methods for assessing drug synergism exist. The method referred to here (REF. 104) is one of the simplest.

**Acknowledgements**

This work was supported by National Institutes of Health grants NS073974 and NS019895.